# Wafer-scale Heterogeneous Integration of Monocrystalline β-Ga$_2$O$_3$ Thin Films on SiC for Thermal Management by Ion-Cutting Technique

Zhe Cheng[1,a)], Fengwen Mu[2,4,a),*], Tiangui You[3,a)], Wenhui Xu[3], Jingjing Shi[1], Michael E. Liao[5], Yekan Wang[5], Kenny Huynh[5], Tadatomo Suga[4], Mark S. Goorsky[5], Xin Ou[3,*], Samuel Graham[1,6,*]

[1] George W. Woodruff School of Mechanical Engineering, Georgia Institute of Technology, Atlanta, Georgia 30332, USA

[2] Kagami Memorial Research Institute for Materials Science and Technology, Waseda University, Shinjuku, Tokyo 169-0051, Japan

[3] State Key Laboratory of Functional Materials for Informatics, Shanghai Institute of Microsystem and Information Technology, Chinese Academy of Sciences, 200050, China

[4] Collaborative Research Center, Meisei University, Hino-shi, Tokyo 191-8506, Japan

[5] Materials Science and Engineering, University of California, Los Angeles, Los Angeles, CA, 90095, United States

[6] School of Materials Science and Engineering, Georgia Institute of Technology, Atlanta, Georgia 30332, USA

[a)] These authors contributed equally

[*] Corresponding authors: mufengwen123@gmail.com; ouxin@mail.sim.ac.cn; sgraham@gatech.edu

# Abstract

The ultra-wide bandgap, high breakdown electric field, and large-area affordable substrates make β-Ga$_2$O$_3$ promising for applications of next-generation power electronics while its thermal conductivity is at least one order of magnitude lower than other wide/ultrawide bandgap semiconductors. To avoid the degradation of device performance and reliability induced by the localized Joule-heating, aggressive thermal management strategies are essential, especially for high-power high-frequency applications. This work reports a scalable thermal management strategy to heterogeneously integrate wafer-scale monocrystalline β-Ga$_2$O$_3$ thin films on high thermal conductivity SiC substrates by ion-cutting technique. The thermal boundary conductance (TBC) of the β-Ga$_2$O$_3$-SiC interfaces and thermal conductivity of the β-Ga$_2$O$_3$ thin films were measured by Time-domain Thermoreflectance (TDTR) to evaluate the effects of interlayer thickness and thermal annealing. Materials characterizations were performed to understand the mechanisms of thermal transport in these structures. The results show that the β-Ga$_2$O$_3$-SiC TBC values increase with decreasing interlayer thickness and the β-Ga$_2$O$_3$ thermal conductivity increases more than twice after annealing at 800 °C due to the removal of implantation-induced strain in the films. A Callaway model is built to understand the measured thermal conductivity. Small spot-to-spot variations of both TBC and Ga$_2$O$_3$ thermal conductivity confirm the uniformity and high-quality of the bonding and exfoliation. Our work paves the way for thermal management of power electronics and β-Ga$_2$O$_3$ related semiconductor devices.

Due to the ultra-wide bandgap (4.8 eV), high breakdown electric field (8 MV/cm), and large-area affordable substrates grown from melt, β-Ga$_2$O$_3$ has attracted great attention for potential applications of next-generation power electronics recently.[1-4] The Baliga figure of merit of β-Ga$_2$O$_3$ is several orders of magnitude higher than that of Si.[3,5,6] However, the thermal conductivity of bulk β-Ga$_2$O$_3$ is at least one order of magnitude lower than those of other wide bandgap semiconductors and is highly anisotropic due to the low (monoclinic) crystal structure.[7-10] Moreover, the thermal conductivity of β-Ga$_2$O$_3$ nanostructures, such as thin films, superlattice, and nanocrystalline films, is further significantly reduced compared with that of the bulk β-Ga$_2$O$_3$.[7,11,12] For high-frequency and high-power applications, the device performance and reliability are limited by the high channel temperature induced by the localized Joule-heating near the gate. Therefore, proper, even aggressive, thermal management is of great importance to avoid device degradation.[4] However, compared with current demonstrations of β-Ga$_2$O$_3$ devices, a disproportionately small amount of work is addressing this sensitive issue.[4]

To extract the heat out of power electronic devices, high thermal conductivity substrates or top metal contacts with high thermal boundary conductance (TBC) near the gate are favorable.[10,13,14] However, direct growth of high-quality β-Ga$_2$O$_3$ on high thermal conductivity substrates such as diamond and SiC is difficult due to the differences in crystal symmetry between β-Ga$_2$O$_3$ and substrates.[12] Recently, devices based on mechanically exfoliated monocrystalline β-Ga$_2$O$_3$ nano-membranes have been fabricated and a record high drain current has been achieved via van der Waals bonding with single crystalline diamond.[15-19] These works demonstrated the great potential of β-Ga$_2$O$_3$ devices. However, these devices have limited capability to scale up to wafer-scale fabrications. Moreover, the van der Waals bonding leads to a low TBC between β-Ga$_2$O$_3$ and the substrates, which hinders the cooling effect of the high thermal conductivity substrates.[11]

In this work, we report a scalable wafer-scale strategy to heterogeneously integrate nanoscale monocrystalline β-Ga$_2$O$_3$ thin films on high thermal conductivity SiC substrates via ion-cutting and surface-activated bonding techniques. Time-domain thermoreflectance (TDTR) is used to measure the thermal conductivity of the β-Ga$_2$O$_3$ thin films and the TBC of the β-Ga$_2$O$_3$-SiC interfaces. Detailed materials characterizations such as a transmission electron microscopy (TEM), High resolution X-ray diffraction (HRXRD), and electron energy loss spectroscopy (EELS) are performed to study the structure-property relation of the β-Ga$_2$O$_3$ thin films and the interfaces. Additionally, we build some thermal models to understand the measured thermal conductivity and TBC.

**Results**

Four samples (10 mm x 8 mm) were prepared in this work, as shown in the schematic diagram of Figure 1. 2-inch bulk ($\bar{2}$01) β-Ga$_2$O$_3$ wafers were implanted with hydrogen ions. A layer of Al$_2$O$_3$ is deposited on the β-Ga$_2$O$_3$ wafers by atomic layer deposition (ALD) before bonding to a 4-inch 4H-SiC wafer at room temperature via the surface activated bonding technique. The Al$_2$O$_3$ layers are used to block the electrical breakdown due to the interface defects in devices while they induce additional thermal resistance. After that, a thin layer of monocrystalline β-Ga$_2$O$_3$ was exfoliated from the bulk β-Ga$_2$O$_3$ wafer by annealing. The β-Ga$_2$O$_3$ wafers can be used for multiple exfoliations, which save materials and lower the fabrication cost. More details about the sample preparation can be found in the Methods section. We name the four samples as Samp1, Samp2,

Samp3, and Samp4. Samp1 and Samp2 have 30-nm-Al$_2$O$_3$ interlayers while Samp3 and Samp4 have10-nm-Al$_2$O$_3$ interlayers. Samp1 and Samp3 are as-bonded while Samp2 and Samp4 were annealed at 800 ºC in N$_2$.

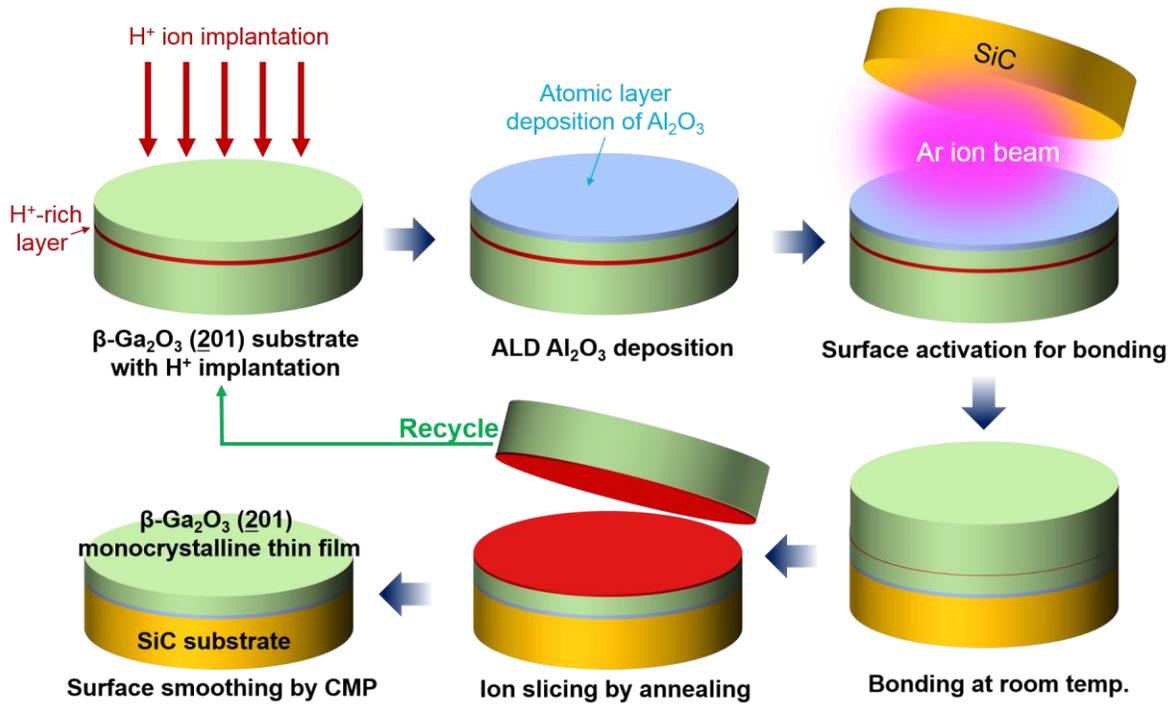

Figure 1. Schematic diagram of sample preparation. A layer of Al$_2$O$_3$ is deposit on the β-Ga$_2$O$_3$ wafer which is implanted by H ions. After that, surface activated bonding is used to bond the Al2O3 coated β-Ga$_2$O$_3$ wafer with a 4H-SiC wafer at room temperature. A thin layer of monocrystalline β-Ga$_2$O$_3$ is transferred on the SiC wafer after heating up to 450 ºC. The surface can be smoothed by chemical mechanical polishing (CMP) and the rest of the β-Ga$_2$O$_3$ wafer can be repeatedly used.

Figure 2 (a) shows a 2-inch ($\bar{2}$01) β-Ga$_2$O$_3$ wafer bonded with a 4-inch 4H-SiC wafer. We can see the excellent quality of the wafer bonding. No voids are observed in the bonded area. Our work brings the bulk-quality monocrystalline ($\bar{2}$01) β-Ga$_2$O$_3$ directly bonded to high thermal conductivity SiC. Figure 2(b) shows the image of the monocrystalline ($\bar{2}$01) β-Ga$_2$O$_3$ transferred onto the 4H-SiC substrate. Only a monocrystalline nanoscale thin film (~140 nm thick) remains on the SiC wafer, which allows for later device fabrication or homo-epitaxial growth of β-Ga$_2$O$_3$ structures on these β-Ga$_2$O$_3$-on-SiC wafers. We did not see any voids on these exfoliated β-Ga$_2$O$_3$ thin films, which confirms the high-quality bonding and exfoliation. Furthermore, the bulk ($\bar{2}$01) β-Ga$_2$O$_3$ wafer can be repeatedly used for multiple exfoliations.

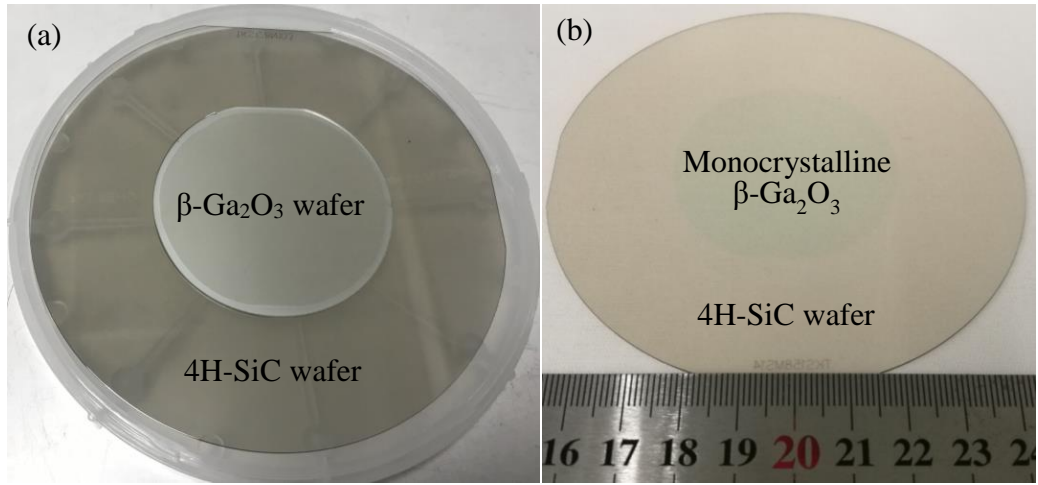

Figure 2. (a) A 2-inch β-Ga$_2$O$_3$ wafer is bonded on a 4-inch 4H-SiC wafer. (b) The monocrystalline β-Ga$_2$O$_3$ thin film is transferred onto the 4H-SiC wafer.

To evaluate the heat dissipation capability of these β-Ga$_2$O$_3$-on-SiC wafers, we measured the thermal boundary conductance (TBC) of the β-Ga$_2$O$_3$-SiC interfaces and the thermal conductivity of the β-Ga$_2$O$_3$ thin films by TDTR. TDTR is an optical pump-probe technique for thermal characterization of both bulk and nanostructured materials.[9,20-22] More details about the TDTR measurements can be found in the Methods section. Here, we will discuss the β-Ga$_2$O$_3$-SiC TBC first. As shown in Figure 3(a), the samples have three layers: Al transducer, β-Ga$_2$O$_3$ layer and SiC substrate. Here, the measured TBC is an effective TBC which includes the thermal resistance of the β-Ga$_2$O$_3$-Al$_2$O$_3$ interface, the Al$_2$O$_3$-interlayer, and the Al$_2$O$_3$-SiC interface. The TBC of the four samples were measured to evaluate the effects of Al$_2$O$_3$ thickness and annealing. Figure 3(b) shows the β-Ga$_2$O$_3$-SiC TBC of Samp1 (30 nm Al$_2$O$_3$ interlayer, as-bonded) and Samp2 (30 nm Al$_2$O$_3$ interlayer, annealed at 800 °C in N$_2$). 14 arbitrarily selected spots were measured on each sample to check the uniformity of the bonding. The TBC variations are less than 6%, which confirms the uniformity of the bonded interfaces. Figure 3(c) shows the measured β-Ga$_2$O$_3$-SiC TBC of Samp3 (10 nm Al$_2$O$_3$ interlayer, as-bonded) and Samp4 (10 nm Al$_2$O$_3$ interlayer, annealed at 800 °C in N$_2$). Uniform TBC is also observed for four different randomly selected spots on the samples. The β-Ga$_2$O$_3$-SiC TBC of Samp3 and Samp4 are larger than those of Samp1 and Samp2 with thicker Al$_2$O$_3$ interlayers. Al$_2$O$_3$ interlayers induce additional thermal resistances at the β-Ga$_2$O$_3$-SiC interfaces so thinner interlayers lead to smaller thermal resistances and larger effective TBC. The TBC of the as-bonded, unannealed samples are slightly larger than those of the annealed interfaces.

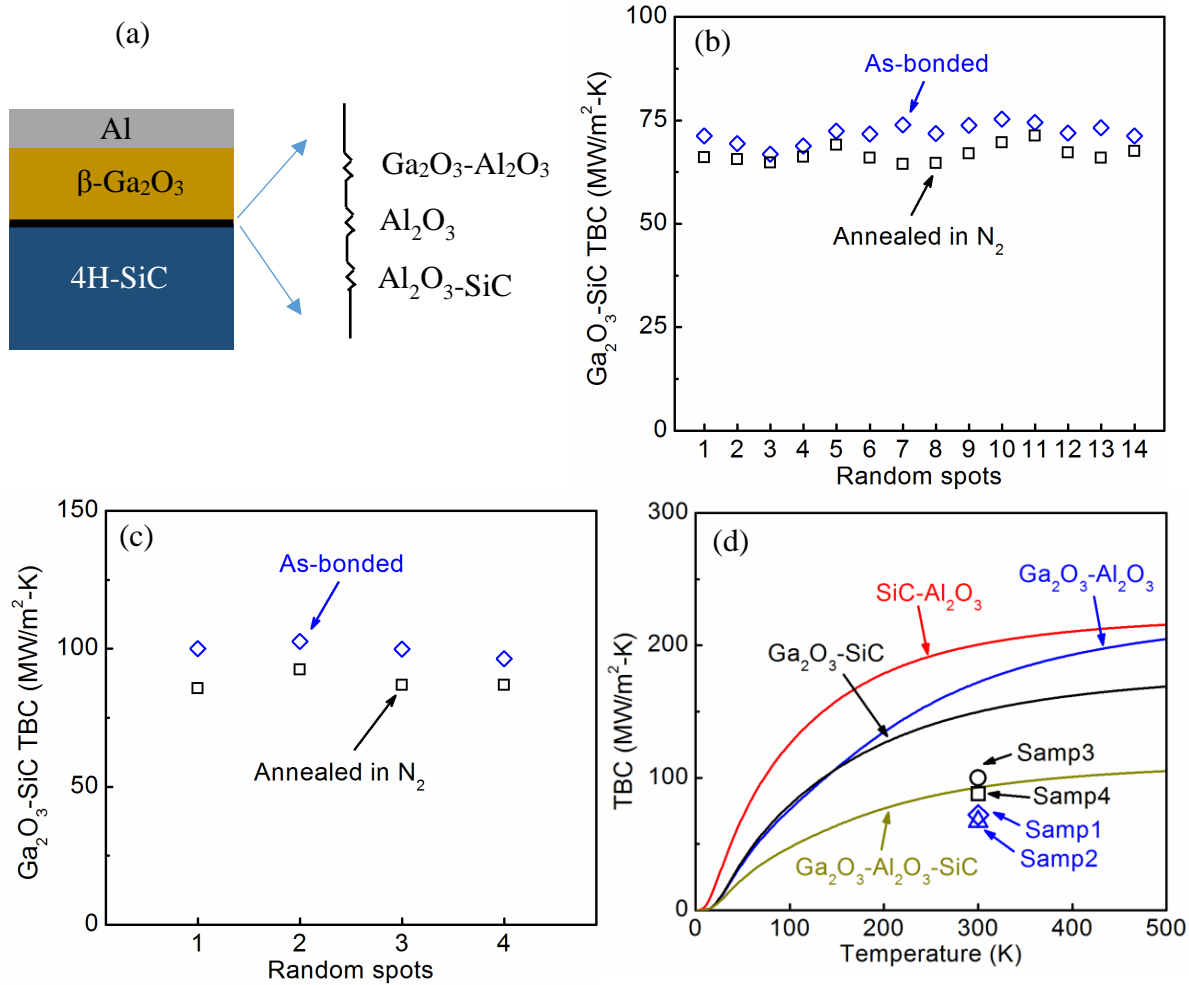

Figure 3. (a) The schematic diagram of the sample structure and thermal resistance. (b) The measured β-$Ga_2O_3$-SiC TBC of Samp1 and Samp2 which have 30 nm $Al_2O_3$ interlayers. 14 different spots were measured to show the uniformity of the samples. (c) The measured β-$Ga_2O_3$-SiC TBC of Samp3 and Samp4 which have 10 nm $Al_2O_3$ interlayers. (d) Comparison of measured TBC and calculated TBC.

Figure 3(d) shows the comparison of the measured TBC of the four samples and calculated temperature dependent TBC of β-$Ga_2O_3$-SiC interfaces, β-$Ga_2O_3$-$Al_2O_3$ interfaces, and $Al_2O_3$-SiC interfaces. Density-functional-theory is used to calculate the phonon dispersion relation of the β-Ga$_2$O$_3$, Al$_2$O$_3$, and 4H-SiC. Diffuse mismatch model (DMM) is used to calculate the phonon transmission and the TBC can be evaluated by the Landauer Formula.[11,23] The detailed derivation and formula used to do this calculation can be found in the Methods section. The β-Ga$_2$O$_3$-SiC and Ga$_2$O$_3$-Al$_2$O$_3$ TBC are not very large due to the small phonon group velocity (6620 m/s) of β-Ga$_2$O$_3$ in the direction perpendicular to ($\bar{2}$01) plane.[11,24] The β-Ga$_2$O$_3$-Al$_2$O$_3$-SiC TBC is an effective TBC which considers the thermal resistance of both the β-Ga$_2$O$_3$-Al$_2$O$_3$ interface and the Al$_2$O$_3$-SiC interface: 1/TBC(β-Ga$_2$O$_3$-Al$_2$O$_3$-SiC)=1/TBC(β-Ga$_2$O$_3$-Al$_2$O$_3$)+1/TBC(Al$_2$O$_3$-SiC). We see good agreement between of the measured TBC and calculated TBC at room temperature. Please note that the DMM-based Landauer formula is only an estimate of the TBC and does not capture exact physics near the interface.

To understand the measured TBC, TEM is used to characterize the interfacial structure. Figure 4(a-b) shows the interfacial structures of Samp1 and Samp2. The β-Ga$_2$O$_3$ is very well-bonded with the SiC wafer without any voids. The β-Ga$_2$O$_3$-Al$_2$O$_3$ interfaces are very smooth and both sides of these interfaces are crystalline materials, while there are amorphous layers at the Al$_2$O$_3$-SiC interfaces. These amorphous layers result from the surface activation by Ar ions during the surface activated bonding process. By comparing Samp1 and Samp2, the annealing process makes the amorphous layers slightly thinner (from ~3.5 nm to ~2 nm), which facilitates thermal transport across interfaces. Also, annealing can possibly improve the crystalline quality of the Al$_2$O$_3$ interlayers, which facilitates thermal transport as well. However, we observed a slightly reduced TBC, especially for Samp3 and Samp4.

To figure out the potential mechanisms of the reduced TBC after annealing (comparison of Samp3 and Samp4), STEM and EELS were performed to study the interfacial structures of Samp3 and Samp4, as shown in Figure 4(c-f). Figure 4 (c-d) show the STEM bright-field (BF) images of the β-$Ga_2O_3$-SiC interfaces of Samp3 and Samp4, respectively. The interfaces consistent of ~9-nm-thick ALD-$Al_2O_3$ layers and ~2-nm amorphous SiC layers. Similar to the amorphous layers in Samp1 and Samp2, these amorphous layers result from the surface activation by the Ar ion beams. The ions impact the SiC surface and break the bonds near the surface, which turns the crystalline SiC into amorphous SiC. Figure 4 (e-f) show the EELS mapping of Ga element distributions of the interface of Samp3 and Samp4. Ga element or $GaO_x$ diffusion into the $Al_2O_3$ interlayers to form AlGaO alloys is observed after annealing in $N_2$ gas at 800 ºC. Alloy structure scatters phonons extensively and reduce thermal conductivity. We attribute this alloy effect as the possible reason for the slightly reduced TBC measured by TDTR.

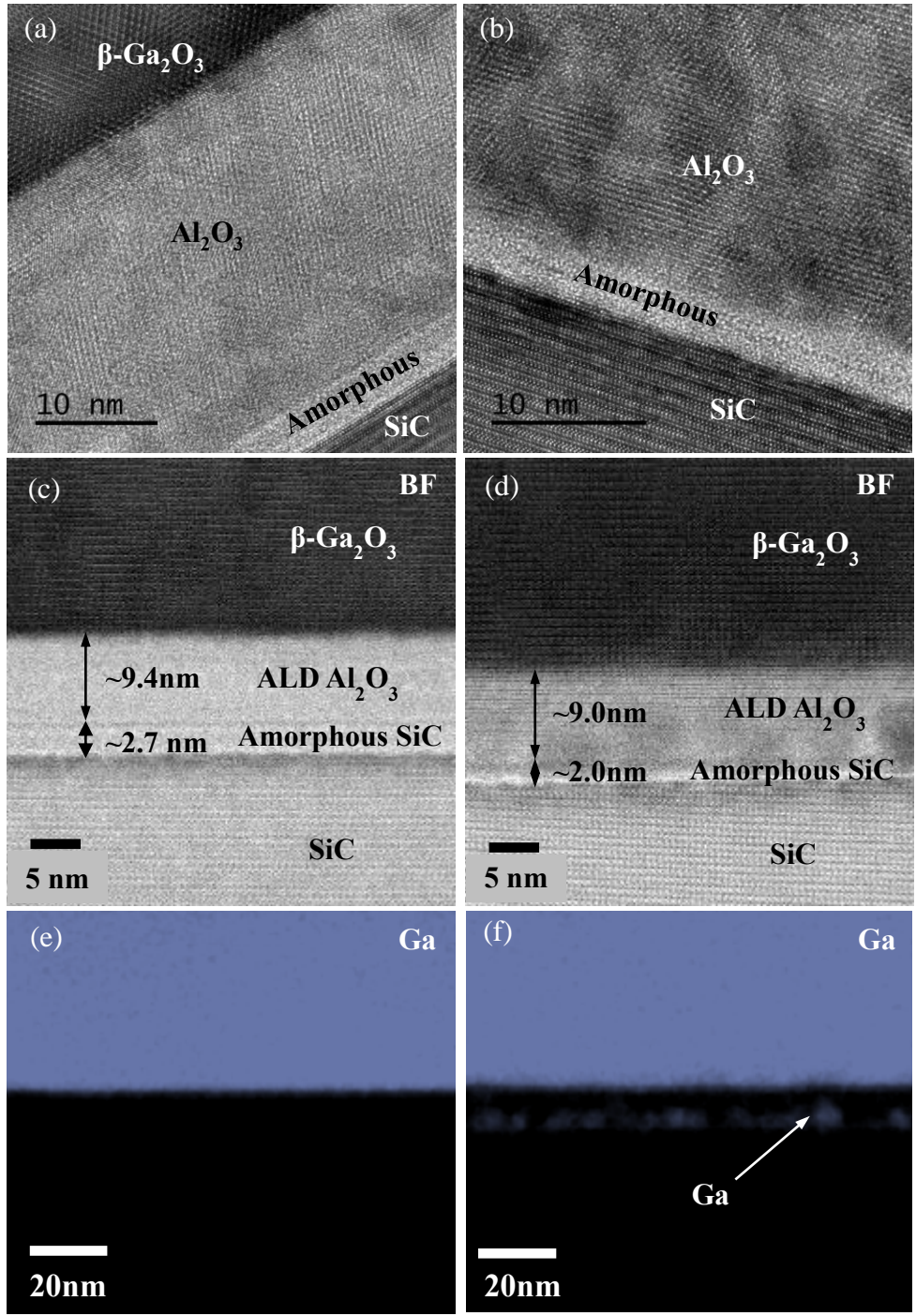

Figure 4. (a) HRTEM images of Samp1 with a 30 nm Al$_2$O$_3$ interlayer (as-bonded). (b) HRTEM images of Samp2 with a 30 nm Al$_2$O$_3$ interlayer (annealed). (c) STEM images of the interface of Samp3 with a 10 nm Al$_2$O$_3$ interlayer (as-bonded). (d) STEM images of the interface of Samp4 with a 10-nm Al$_2$O$_3$ interlayer (annealed). (e) Ga element EELS map of the interface of Samp3.

(f) Ga EELS map of the interface of Samp4. Ga diffusion is observed after annealing in $N_2$ gas at 800 °C.

The thermal conductivity of the monocrystalline nanoscale β-$Ga_2O_3$ thin films were also measured by TDTR, as shown in Figure 5. Figure 5(a) shows the measured thermal conductivity of β-$Ga_2O_3$ thin films of Samp1 and Samp2 while Figure 5(b) shows those of Samp3 and Samp4. 14 random spots were measured to check the uniformity of the β-$Ga_2O_3$ thin films. The small variation of the measured thermal conductivity confirms the uniformity of these thin films. The measured thermal conductivity of the monocrystalline ($\underline{2}$01) β-$Ga_2O_3$ thin films (2.9 W/m-K) is more than four times lower than the bulk counterpart (13.3 W/m-K).[24] The significant reduction in thermal conductivity is due to the film boundary scattering and the implantation-induced strain in the thin films, which will be discussed in detail later.[25] The thermal conductivity of the β-$Ga_2O_3$ thin films increases more than twice after annealing in $N_2$. Annealing reduces the implant-induced strain in the thin films. As shown in Figure 5(b), a similar increase in the thermal conductivity of the β-$Ga_2O_3$ thin films was observed in Samp3 and Samp4.

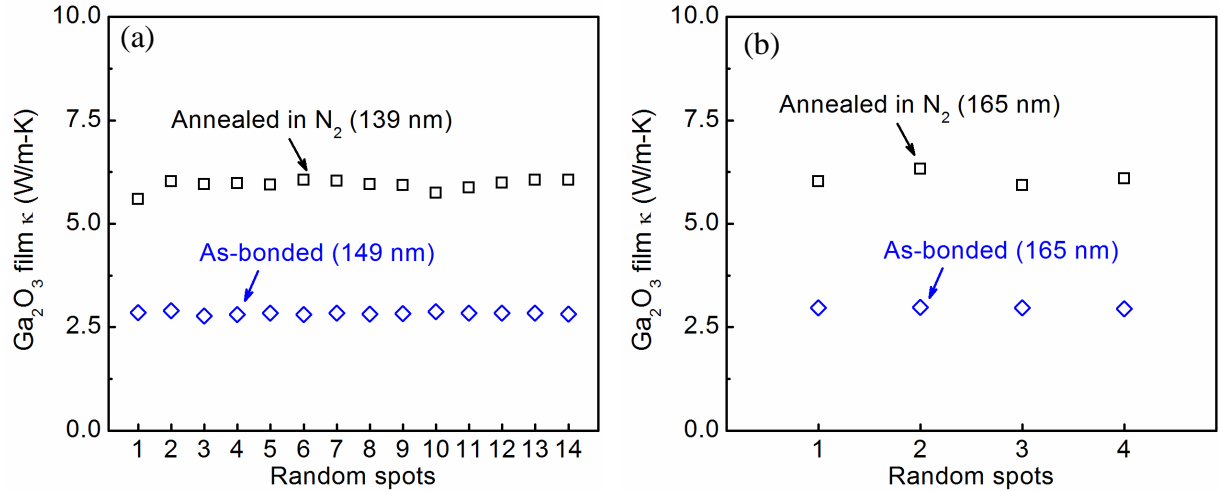

figure 5. (a) The measured thermal conductivity of β-Ga$_2$O$_3$ thin films of Samp1 and Samp2. (b) The measured thermal conductivity of β-Ga$_2$O$_3$ thin films of Samp3 and Samp4.

To evaluate the strain variation before and after annealing in N$_2$, we performed triple-axis XRD measurements on Samp1 and Samp2, as shown in Figure 6(a-b). For the as-bonded sample, the strain from implantation is not completely removed while all strain is removed for the annealed samples. This is similar to other implanted materials.[25,26] Figure 6(a) is the triple-axis diffraction and the peaks are relative to the (0004) 4H-SiC. The peaks near ~ -30,000 arcsec are from the β-Ga$_2$O$_3$ layers; this large offset shows the angular difference between the ($\underline{2}$01) β-Ga$_2$O$_3$ peak and the (0004) 4H-SiC peak. Samp1 shows an asymmetric peak and is shifted to a lower angle (larger out-of-plane lattice parameter as expected for implanted material) from the expected lattice parameter of β-Ga$_2$O$_3$. The fringe period corresponds to a thickness of ~140 nm, which agrees with the thicknesses measured by TEM. Samp2 shifts to the smaller angle expected for ($\underline{2}$01) β-Ga$_2$O$_3$ and shows a symmetric peak with a fringe period that also corresponds to ~140 nm.

Figure 6(b) shows the rocking curves of Samp1 (as-bonded) and Samp2 (annealed at 800 °C in $N_2$). The FWHM of Samp1 is 170" while that of Samp2 is 130". Annealing improves the crystallinity but the broader shoulders of the annealed sample shows other defects may evolve during the annealing. Figure 6(c) shows the comparison of calculated and measured thermal conductivity. The four samples studied in this work were heated to 450 °C during the exfoliation process. Samp2 and Samp4 were annealed at 800 °C in $N_2$. The 450 °C heating drives the H ions out of β-$Ga_2O_3$ films while the high temperature annealing (800 °C) can improve the quality of the films by reducing the amounts of defects (vacancies) in the films. Phonons scatter with defects due to mass difference, and the local strain which induces changes in local bond stiffness, lattice constant, and Gruneisen parameter.[27] If we do not consider mass difference, then the local strain field is the primary source for phonon-defect scatterings, similar to other implanted materials.[27] The strain-induced phonon-defect scatterings reduce the thermal conductivity of these β-$Ga_2O_3$ thin films. The line labeled with "No strain" in Figure 6(c) shows the thickness dependent thermal conductivity which only considers the phonon-phonon scatterings and phonon-boundary scatterings. The measured thermal conductivity is slightly lower than the calculated thermal conductivity because of the remaining small amount of structural imperfection, which was confirmed by the shoulder in the XRD curve. The line labeled with "Strain" in Figure 6(c) shows the thickness dependent thermal conductivity which adds additional phonon-defect scatterings. By adding the phonon-defect scatterings into the calculation, we can match the measured thermal conductivity of the as-bonded samples to obtain the scattering cross-section ($\Gamma_i$) as 0.3. The phonon scattering mechanisms are shown in Figure 6(d). The strain filed induced phonon-defect scattering exists in the films while the phonon-boundary scatterings happen at the film boundaries. Both the phonon-boundary scatterings and phonon-defect scatterings reduce thermal conductivity of these thin films. The phonon-phonon scatterings are not included in the figure because they exist in both the films with strain and films without strains.

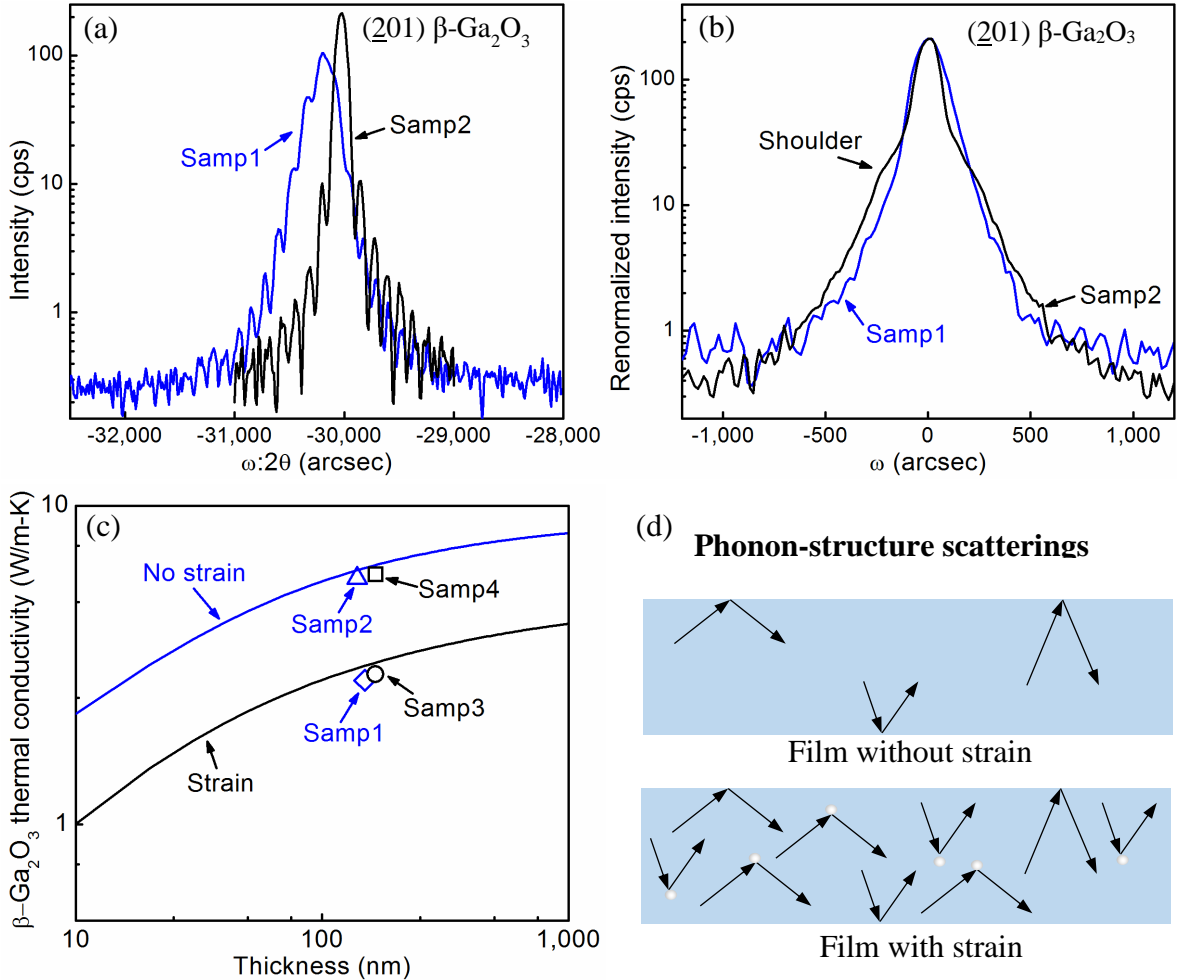

Figure 6. (a) Triple-axis (ω:2θ) scans of Samp1 and Samp2. (b) Triple-axis rocking curves of Samp1 and Samp2. The FWHM of Samp1 is 170" while that of Samp2 is 130". (c) Comparison of calculated and measured thermal conductivity of β-Ga₂O₃ thin films. (d) Phonon scatterings with film boundaries and defects. The strain in the as-bonded samples induces phonon-defect scattering and correspondingly reduce thermal conductivity.

## Conclusions

This work reports a scalable thermal management strategy to heterogeneously integrate wafer-scale monocrystalline β-Ga$_2$O$_3$ thin films on high thermal conductivity 4H-SiC substrates by ion-cutting and surface activated bonding techniques. We observe medium TBC values (67-100 MW/m$^2$-K) of the β-Ga$_2$O$_3$-SiC interfaces. The TBC values increase about 39% and 32% when decreasing the Al$_2$O$_3$ interlayer thickness from 30 nm to 10 nm for the as-bonded and annealed interfaces, respectively. The thermal annealing reduces the β-Ga$_2$O$_3$-SiC TBC slightly possibly due to Ga element or GaO$_x$ diffusion into the Al$_2$O$_3$ interlayer. The β-Ga$_2$O$_3$ thermal conductivity increases by a factor of two after annealing in N$_2$ at 800 °C due to the removal of implantation-induced strain in the films. Some theoretical models were used to understand the measured TBC and thermal conductivity. Fourteen different spots were measured to evaluate the uniformity. The small spot-spot variations of both TBC and β-Ga$_2$O$_3$ thermal conductivity confirm the uniformity and effectiveness of the bonding, exfoliation, and annealing. Our work paves the way for thermal management of power electronics and β-Ga$_2$O$_3$ related semiconductor devices.

## Methods

**Sample Preparation.** The on-axis (0001) 4-inch 4H-SiC wafers with a thickness of ~500 μm were purchased from SICC Co., Ltd. The ($\bar{2}$01) 2-inch β-Ga$_2$O$_3$ wafers with a thickness ~680 μm were purchased from Novel Crystal Technology Inc. The Si-face of the 4H-SiC was used as the bonding surface. All of the bonding surfaces were polished by chemical mechanical polishing (CMP). The root-mean-square (RMS) surface roughness of the SiC and β-Ga$_2$O$_3$ surfaces are ~0.30 nm and ~0.27 nm, respectively.

The thin film exfoliation can be achieved by either hydrogen or helium ion implantation.[28] In the present study, the β-Ga$_2$O$_3$ wafers were implanted by hydrogen ions (H$^+$) with a 7-degree tilt to minimize the ion channeling effect.[29] The implanted H$^+$ ions accumulate in the β-Ga$_2$O$_3$ at 200-400 nm beneath the surface. After the ion implantation, an Al$_2$O$_3$ film was deposited on the Ga$_2$O$_3$ wafers by plasma-enhanced atom-layer deposition (PEALD) to avoid the surface damage layer during the following surface activation process for bonding and also blocking possible electrical breakdown due to the defective interfaces. Trimethylaluminum (TMA) and O$_2$ plasma were used as ALD precursors. Ar gas flow served as both carriers and purging gas. The O$_2$ plasma power was 100 W. The total flow rate of the Ar was 200 sccm (standard cubic centimeters per minute). The Al$_2$O$_3$ thin films were grown at 200 ºC. The growth rate per cycle (GPC) was typically 1.3 Å/cycle for Al$_2$O$_3$. Al$_2$O$_3$ films with two kinds of thickness (30 nm for Samp1 and Samp2 and 10 nm for Samp3 and Samp4) were deposited on the implanted β-Ga$_2$O$_3$ wafers. After that, the two wafers were bonded together at room temperature by the surface-activated bonding (SAB) technique. The SAB machine consists of a load-lock chamber and a processing-bonding chamber. A silicon source is placed in the ion source to generate a mixture of Si ions and Ar ions. The small amount of Si ions added into the Ar ion beam was aimed to reduce the carbonation of SiC surface during the ion beam bombardment. The Ar ions activate the SiC and Al$_2$O$_3$ (on the β-Ga$_2$O$_3$) surfaces to create dangling bonds. After the surface activation, the samples were bonded directly at room temperature. The detailed bonding process can be found in the literature.[30]

By heating the bonded wafers at 450 °C, the implanted H$^+$ ions accumulate near the projected range of the ion implantation to induce the exfoliation of a β-Ga$_2$O$_3$ thin film.[26,29,31,32] After that, a β-Ga$_2$O$_3$ thin film with a thickness of less than 400 nm was split from the bulk β-Ga$_2$O$_3$ wafer and transferred to SiC substrate. The exfoliated β-Ga$_2$O$_3$ thin film was thinned by CMP to obtain a good TDTR sensitivity of the buried β-Ga$_2$O$_3$-SiC interface (100-200 nm thick). The post-exfoliated β-Ga$_2$O$_3$ bulk wafer can be repeatedly used for multiple exfoliations after polishing. The transferred 2-inch β-Ga$_2$O$_3$-on-SiC wafers were diced into 10 mm × 8 mm chips. Some chips were annealed in flowing N$_2$ gas. The temperature was ramped to 800 °C with a rate of 5 °C/min and then held at 800 °C for 30 min. Finally, a ~100-nm-thick Al TDTR transducer layer was deposited on the surface of all samples.

**TDTR Measurements**. TDTR is a femtosecond-laser-based optical pump-probe technique to measure thermal properties of both bulk and nanostructured materials.[9,20] A modulated pump beam heats the sample surface periodically while a delayed probe beam detects the surface temperature variation via thermoreflectance.[22] The variation of surface temperature is fitted with an analytical heat transfer solution of the samples to infer the unknown thermal parameters.[10,22] The measurements in this work are similar to those in references.[10,11,33,34]

A 10 X objective (pump radius 10.1 μm and probe radius 5.8 μm) is used for the measurements with modulation frequency of 2.2 MHz. The low modulation frequency is to obtain deep thermal penetration depth to get enough sensitivity for the buried β-Ga$_2$O$_3$- SiC interfaces. More information about TDTR sensitivity can be found in the Supplementary Information. A bare SiC wafer was measured first to obtain the SiC thermal conductivity (337 W/m-K), which was used as a known parameter in the data fitting of measurements on β-Ga$_2$O$_3$ on SiC samples. The Al and β-Ga$_2$O$_3$ thicknesses were measured by the picosecond acoustic technique and confirmed with TEM results. The thermal conductivity of Al is obtained by measuring the electrical conductivity and applying Wiedemann-Franz law. The heat capacity of the Al transducer, the β-Ga$_2$O$_3$ thin film, and the SiC substrates are from literature.[22,24,33,35]

**Materials Characterization.** Cross-section TEM samples were prepared with a FEI Helios dual beam focused ion beam (FIB) system. The interfacial structures were characterized by a HR-STEM (Probe-corrected FEI Titan) and the interfacial composition was measured by EELS (Gatan Enfinium). The observation in this study is along <11-20> axis of SiC and <010> axis of β-Ga$_2$O$_3$. Triple-axis x-ray diffraction measurements were performed on a Bruker-JV D1 high-resolution X-ray diffractometer with incident beam conditioning that includes a parallel beam optical element and a (110) channel-cut silicon crystal that produces a monochromatic, highly collimated Cu Kα$_1$ beam. The scattered beam optics also included a (110) channel-cut silicon crystal. Both longitudinal (Δ (ω:2θ)) scans and rocking curves (Δω) were recorded under these triple axis HRXRD conditions.

**Theoretical Modelings.** A Landauer approach with DMM is used to calcuate the β-Ga$_2$O$_3$-Al$_2$O$_3$, Al$_2$O$_3$-SiC, and β-Ga$_2$O$_3$-SiC TBC. Because the transmission function from DMM does not depend on the angle of incidence, the Landauer formula is:

$$G = \sum_p \frac{1}{4} \int D_1(\omega) \frac{df_{BE}}{dT} \hbar \omega v_1(\omega) \tau_{12}(\omega) d\omega. \qquad (1)$$

where $D$ is the phonon density of states, $f_{BE}$ is the Bose-Einstein distribution function, $\hbar$ is the reduced Planck constant, $\omega$ is the phonon angular frequency, $v$ is the phonon group velocity of material 1, $\tau_{12}$ is the transmission coefficient from material 1 to 2, and the sum is over all incident phonon modes. The expression of the transmission function from DMM[23] is

19... 

$$\tau_{12}(\omega) = \frac{\sum_p M_2(\omega)}{\sum_p M_1(\omega) + \sum_p M_2(\omega)}, \qquad (2)$$

where $M$ is the phonon number of modes. The phonon properties of diamond are obtained from first principles calculation with VASP.

A Callaway model is used to understand the measured β-Ga$_2$O$_3$ thermal conductivity by taking account of the film boundary scattering and strain effect in the films.[36-38] The phonon dispersion relation of β-Ga$_2$O$_3$ is calculated by density-functional-theory (DFT). All the phonon polarizations are considered (acoustic and optical, totally 30 branches for β-Ga$_2$O$_3$). The Callaway model gives an expression of thermal conductivity $k$ as an integration in frequency space:

$$k = \frac{1}{3} \sum_p \int_0^{\omega_{\text{cut-off}}} \hbar \omega D_\lambda \frac{df_{BE}}{dT} v_\lambda^2 \tau_{C,\lambda}, \qquad (3)$$

where the 1/3 comes from the isotropic assumption, $\sum_p$ is over all phonon polarizations, $v_\lambda$ is the modal phonon group velocity, $\tau_{C,\lambda}$ is the modal combined relaxation time of phonon mode $\lambda$, $D_\lambda$ is the modal phonon density of states, and $f_{BE}$ is the Bose-Einstein distribution function. Three phonon scattering mechanisms are included in our calculation: phonon-phonon scatterings, phonon-boundary scatterings, and phonon-defect scatterings. The effective relaxation time $\tau_{C,\lambda}$ of each phonon mode can be obtained from the Matthiessen's rule[38]:

$$\tau_{C,\lambda} = \left( \frac{1}{\tau_U} + \frac{1}{\tau_D} + \frac{1}{\tau_B} \right)^{-1} \qquad (4)$$

where $\tau_U$, $\tau_D$, and $\tau_B$ are the relaxation time of Umklapp phonon-phonon scatterings, phonon-defect scatterings, and phonon-boundary scatterings, respectively. The scattering rate can be shown below.[27,39]

$$\frac{1}{\tau_U} = BT\omega^2 e^{-\frac{C}{T}}, \qquad (5)$$

$$\frac{1}{\tau_D} = \frac{V\omega^4}{4\pi v^3}\Gamma_i, \qquad (6)$$

$$\frac{1}{\tau_B} = v/d, \qquad (7)$$

where $B$ and $C$ are fitting parameters, $V$ is the atomic volume of the β-Ga$_2$O$_3$, $\Gamma_i$ is the parameter to describe the strength of phonon-defect scattering of defect $i$, and d is the thickness of the β-Ga$_2$O$_3$ film. When calculating the scattering rates of different mechanisms, the $B$ (2.3 ×10$^{-18}$ s/K) and $C$ (130 K) are obtained by fitting with the previous experimental measurements.[8,24]

**Competing interests:** The authors claim no competing financial interests.

**Author contributions**: Z. C., J. S., and S. G. did the thermal measurements and calculations. F. M., T. Y., W. X., T. S., and X. O. fabricated the samples and finished part of the material characterizations. M. E. L., Y. W., K. H., and M. S. G. finished the XRD measurements and part of the TEM study. Z. C. wrote the manuscript with inputs from all other authors.

**Acknowledgements**: Z. C., J. S., M. E. L., Y. W., K. H., M. S. G., and S. G. would like to acknowledge the financial support from U.S. Office of Naval Research under a MURI program (Grant No. N00014-18-1-2429). Z. C. J. S., and S. G. acknowledge the funding support from U.S. Air Force Office of Scientific Research under a MURI program (Grant No. FA9550-18-1-0479). F.M., and T.S. would like to acknowledge the financial support from Japan JSPS KAKENHI Grant Number 19K15298 and the Precise Measurement Technology Promotion Foundation. T. Y., W. X., and X. O. was supported by National Natural Science Foundation of China (No. 61874128, 61851406, and 11705262), Frontier Science Key Program of CAS (No. QYZDY-SSW-JSC032), and Shanghai Science and Technology Innovation Action Plan Program (No. 17511106202).

# References

1   Liu, Y. *et al.* van der Waals Integrated Devices Based on Nanomembranes of 3D Materials. *Nano Letters* **20**, 1410-1416 (2020).

2   Higashiwaki, M. & Jessen, G. H.    (AIP Publishing, 2018).

3   Pearton, S. *et al.* A review of Ga2O3 materials, processing, and devices. *Applied Physics Reviews* **5**, 011301 (2018).

4   Reese, S. B., Remo, T., Green, J. & Zakutayev, A. How Much Will Gallium Oxide Power Electronics Cost? *Joule* (2019).

5   Higashiwaki, M., Sasaki, K., Kuramata, A., Masui, T. & Yamakoshi, S. Gallium oxide (Ga2O3) metal-semiconductor field-effect transistors on single-crystal β-Ga2O3 (010) substrates. *Applied Physics Letters* **100**, 013504 (2012).

6   Tsao, J. *et al.* Ultrawide-Bandgap Semiconductors: Research Opportunities and Challenges. *Advanced Electronic Materials* **4**, 1600501 (2018).

7   Cheng, Z. *et al.* Significantly Reduced Thermal Conductivity in Beta-(Al0. 1Ga0. 9) 2O3/Ga2O3 Superlattices. *Applied Physics Letter* **115** (2019).

8   Jiang, P., Qian, X., Li, X. & Yang, R. Three-dimensional anisotropic thermal conductivity tensor of single crystalline β-Ga2O3. *Applied Physics Letters* **113**, 232105 (2018).

9   Cheng, Z. *et al.* Experimental observation of high intrinsic thermal conductivity of AlN. *Physical Review Materials* **4**, 044602 (2020).

10  Cheng, Z., Mu, F., Yates, L., Suga, T. & Graham, S. Interfacial Thermal Conductance across Room-Temperature-Bonded GaN/Diamond Interfaces for GaN-on-Diamond Devices. *ACS Applied Materials & Interfaces* **12**, 8376-8384 (2020).

11 Cheng, Z. *et al.* Thermal conductance across β-Ga2O3-diamond van der Waals heterogeneous interfaces. *APL Materials* **7**, 031118 (2019).

12 Cheng, Z. *et al.* Integration of polycrystalline Ga2O3 on diamond for thermal management. *Applied Physics Letters* **116**, 062105 (2020).

13 Mu, F. *et al.* High Thermal Boundary Conductance across Bonded Heterogeneous GaN–SiC Interfaces. *ACS applied materials & interfaces* **11**, 33428-33434 (2019).

14 Aller, H. T. *et al.* Chemical Reactions Impede Thermal Transport Across Metal/β-Ga2O3 Interfaces. *Nano letters* **19**, 8533-8538 (2019).

15 Hwang, W. S. *et al.* High-voltage field effect transistors with wide-bandgap β-Ga2O3 nanomembranes. *Applied Physics Letters* **104**, 203111 (2014).

16 Ahn, S. *et al.* Effect of front and back gates on β-Ga2O3 nano-belt field-effect transistors. *Applied Physics Letters* **109**, 062102 (2016).

17 Zhou, H., Maize, K., Qiu, G., Shakouri, A. & Ye, P. D. β-Ga2O3 on insulator field-effect transistors with drain currents exceeding 1.5 A/mm and their self-heating effect. *Applied Physics Letters* **111**, 092102 (2017).

18 Kim, J., Mastro, M. A., Tadjer, M. J. & Kim, J. Heterostructure WSe2− Ga2O3 Junction Field-Effect Transistor for Low-Dimensional High-Power Electronics. *ACS applied materials & interfaces* **10**, 29724-29729 (2018).

19 Noh, J., Si, M., Zhou, H., Tadjer, M. J. & Peide, D. Y. in *2018 76th Device Research Conference (DRC).* 1-2 (IEEE).

20 Cahill, D. G. Analysis of heat flow in layered structures for time-domain thermoreflectance. *Review of scientific instruments* **75**, 5119-5122 (2004).

21  Cheng, Z. *et al.* Probing Growth-Induced Anisotropic Thermal Transport in High-Quality CVD Diamond Membranes by Multi-frequency and Multi-spot-size Time-Domain Thermoreflectance. *ACS applied materials & interfaces* (2018).

22  Cheng, Z. *et al.* Tunable Thermal Energy Transport across Diamond Membranes and Diamond-Si Interfaces by Nanoscale Graphoepitaxy. *ACS applied materials & interfaces* (2019).

23  Fisher, T. S. *Thermal energy at the nanoscale*. Vol. 3 (World Scientific Publishing Company, 2013).

24  Guo, Z. *et al.* Anisotropic thermal conductivity in single crystal β-gallium oxide. *Applied Physics Letters* **106**, 111909 (2015).

25  Scott, E. A. *et al.* Orders of magnitude reduction in the thermal conductivity of polycrystalline diamond through carbon, nitrogen, and oxygen ion implantation. *Carbon* **157**, 97-105 (2020).

26  Miclaus, C. & Goorsky, M. Strain evolution in hydrogen-implanted silicon. *Journal of Physics D: Applied Physics* **36**, A177 (2003).

27  Scott, E. A. *et al.* Phonon scattering effects from point and extended defects on thermal conductivity studied via ion irradiation of crystals with self-impurities. *Physical Review Materials* **2**, 095001 (2018).

28  Liao, M. E., Wang, Y., Bai, T. & Goorsky, M. S. Exfoliation of β-Ga2O3 Along a Non-Cleavage Plane Using Helium Ion Implantation. *ECS Journal of Solid State Science and Technology* **8**, P673-P676 (2019).

29  Xu, W. *et al.* in *2019 IEEE International Electron Devices Meeting (IEDM)*.  12.15. 11-12.15. 14 (IEEE).

30  Xu, Y. *et al.* Direct wafer bonding of Ga2O3–SiC at room temperature. *Ceramics International* **45**, 6552-6555 (2019).

31  Hayashi, S., Bruno, D. & Goorsky, M. Temperature dependence of hydrogen-induced exfoliation of InP. *Applied Physics Letters* **85**, 236-238 (2004).

32  Ferain, I., Byun, K. Y., Colinge, C., Brightup, S. & Goorsky, M. Low temperature exfoliation process in hydrogen-implanted germanium layers. *Journal of Applied Physics* **107**, 054315 (2010).

33  Mu, F. *et al.* High Thermal Boundary Conductance across Bonded Heterogeneous GaN-SiC Interfaces. *ACS Applied Materials & Interfaces* **11**, 7 (2019).

34  Gaskins, J. T. *et al.* Thermal Boundary Conductance Across Heteroepitaxial ZnO/GaN Interfaces: Assessment of the Phonon Gas Model. *Nano letters* **18**, 7469-7477 (2018).

35  Zheng, Q. *et al.* Thermal conductivity of GaN, GaN 71, and SiC from 150 K to 850 K. *Physical Review Materials* **3**, 014601 (2019).

36  Callaway, J. Model for lattice thermal conductivity at low temperatures. *Physical Review* **113**, 1046 (1959).

37  Mingo, N. Calculation of Si nanowire thermal conductivity using complete phonon dispersion relations. *Physical Review B* **68**, 113308 (2003).

38  Feng, T. & Ruan, X. Quantum mechanical prediction of four-phonon scattering rates and reduced thermal conductivity of solids. *Physical Review B* **93**, 045202 (2016).

39  Klemens, P. The scattering of low-frequency lattice waves by static imperfections. *Proceedings of the Physical Society. Section A* **68**, 1113 (1955).